\documentclass[a4paper,11pt]{article}

\usepackage[utf8]{inputenc}  
\usepackage[T1,T2A]{fontenc} 
\usepackage[russian,english]{babel}

\usepackage{geometry}
\usepackage{microtype}       
\usepackage{datetime}        
\usepackage{fancyhdr}
\usepackage{titlesec}

\usepackage[table,dvipsnames]{xcolor} 
\definecolor{linkblue}{RGB}{0, 0, 255} 
\usepackage{graphicx}
\usepackage[export]{adjustbox} 
\graphicspath{{./images/}}     

\usepackage{amsmath, amssymb, amsfonts}
\usepackage{mathrsfs}
\usepackage{stmaryrd}
\usepackage{mhchem}

\usepackage{tikz}
\usepackage{pgfplots}
\pgfplotsset{compat=1.17}
\usepgfplotslibrary{fillbetween}
\usetikzlibrary{automata, positioning, calc, patterns, fillbetween}

\makeatletter
\pgfmathdeclarefunction{erf}{1}{%
  \pgfmathparse{(1 - exp(-#1^2 * (4/pi + #1^2))) * sign(#1)}%
}
\pgfmathdeclarefunction{gauss}{3}{%
  \pgfmathparse{1/(#3*sqrt(2*pi))*exp(-((#1-#2)^2)/(2*#3^2))}%
}
\makeatother

\usepackage{booktabs}
\usepackage{tabularx}
\usepackage{multirow}
\usepackage{array}
\usepackage{makecell}
\usepackage{subcaption}

\usepackage{amsthm}

\numberwithin{theorem}{section}

\usepackage{enumitem}
\setlist[itemize]{label=--}

\usepackage{lscape}
\usepackage{float}

\usepackage[font=small, labelfont=bf]{caption}

\usepackage[colorlinks=true, linkcolor=linkblue, citecolor=linkblue, urlcolor=linkblue]{hyperref}
\usepackage{cleveref}
\crefname{figure}{Fig.}{Figs.}
\crefname{table}{Tab.}{Tabs.}
\crefname{equation}{Eq.}{Eqs.}
\crefname{section}{Sec.}{Secs.}
\Crefname{section}{Sec.}{Secs.}

\usepackage[style=numeric, maxcitenames=2, date=year, doi=false, isbn=false, url=false, eprint=false, giveninits=true]{biblatex}
\DeclareNameAlias{author}{last-first}
\DeclareFieldFormat[article]{title}{#1}
\DeclareFieldFormat[article]{journaltitle}{\textit{#1}}
\DeclareFieldFormat[article]{volume}{\textbf{#1}}
\DeclareFieldFormat[article]{number}{(#1)}
\DeclareFieldFormat[article]{pages}{#1}
\renewbibmacro{in:}{}

\usepackage{gitinfo2}

\usepackage[percent]{overpic}

\begin{document}
\title{Quality Control and Structural Reliability -- A Unified Framework for Integrating Conformity Assessment and Partial Safety Factors}

\author{
    Tammam Bakeer\thanks{Dr.-Ing. habil. Tammam Bakeer, E-Mail: \href{mailto:mail@bakeer.de}{mail@bakeer.de}},  
    Wolfram Jäger\thanks{Prof. Dr.-Ing. Wolfram Jäger, E-Mail: \href{mailto:w.jaeger@jaeger-ingenieure.de}{w.jaeger@jaeger-ingenieure.de}} 
}

\maketitle

\begin{abstract}
    Ensuring structural reliability remains a core concern in civil engineering, yet the quantitative effects of quality control measures on material variability and safety margins are not fully understood, especially for materials other than reinforced concrete. This study addresses this gap by presenting a probabilistic framework that integrates Bayesian updating, acceptance sampling, and operating characteristic (OC) curves to model conformity assessment as a probabilistic filter. In doing so, it refines prior distributions of key material and execution parameters based on quality control outcomes, linking reductions in the coefficient of variation directly to adjustments in partial safety factors.

    Applying the framework to a masonry wall example demonstrates how systematic quality control efforts, particularly those targeting parameters with higher importance—such as masonry unit strength and execution quality—produce substantial gains in structural reliability. The analysis shows that combined quality control measures can lower the partial safety factor from a baseline of 1.5 to about 1.38, corresponding to an improvement factor of roughly 1.09 and material savings of approximately 8\%. Conversely, controlling parameters with negligible influence, such as mortar properties, provides limited benefit.
    
    These findings encourage focusing quality control resources on the most influential parameters and integrating results into semi-probabilistic design methods. By offering a transparent, standards-compatible approach, the framework supports the refinement of design guidelines, promotes more efficient resource allocation, and enhances overall structural safety in the built environment.

\end{abstract}

\vspace{1cm} 

\newpage
\tableofcontents


\section{Introduction}

Ensuring the safety and reliability of load-bearing structures is a cornerstone of civil engineering, where structural failures can lead to severe economic, societal, and environmental consequences. The performance of structures depends heavily on the quality of construction materials, execution processes, and the accuracy of design models. Variability in these parameters—whether due to inadequate quality control, material inconsistencies, or imperfections in execution—poses significant challenges to achieving target reliability levels while maintaining cost-efficiency. As modern construction grows in complexity and scale, the need for robust methodologies to systematically manage and quantify uncertainties has never been greater.

Within the European Union, the \textit{Construction Products Regulation} (CPR) \cite{RegulationEUNo2011} establishes a harmonized legal framework for ensuring the performance consistency of construction products. Central to this regulation is the \textit{Assessment and Verification of Constancy of Performance} (AVCP) system, which mandates conformity assessment procedures to identify and reject non-compliant materials or components. This filtering process inherently improves material quality by refining the statistical distributions of critical parameters, such as material strength or initial imperfections. However, despite its widespread implementation, the quantification of conformity assessment’s impact on structural reliability remains inadequately addressed.

Existing research predominantly focuses on reliability-based design methods and the calibration of partial safety factors to ensure structural safety. However, much of this work provides only qualitative insights into the effects of quality control on reducing variability and enhancing reliability. Furthermore, the practical integration of conformity assessment into semi-probabilistic design frameworks has yet to be systematically explored. Recent advances in statistical modeling, particularly \textit{Bayesian updating}, \textit{acceptance sampling}, and \textit{operating characteristic} (OC) curves, provide a rigorous foundation for linking quality control processes to quantifiable improvements in structural reliability. By combining these tools, it becomes possible to evaluate how reductions in material and execution variability directly influence partial safety factors and resource efficiency.

This study introduces a probabilistic framework that integrates conformity assessment into reliability-based design methodologies. By treating quality control as a probabilistic filtering mechanism, the proposed approach employs Bayesian updating and OC curves to quantify reductions in parameter variability and their effect on partial safety factors. A detailed application to a vertically loaded masonry wall serves as a practical case study, illustrating how systematic quality control measures enhance structural performance while achieving cost-effective designs.

The paper is structured as follows:  

\cref{sec:02} reviews the state of the art in reliability assessment and conformity control, identifying existing knowledge gaps.  
\cref{sec:03} introduces the theoretical basis for treating conformity assessment as a filtering process, incorporating Bayesian methods and OC curves.  
\cref{sec:04} outlines the approach for coupling quality control measures with the calibration of partial safety factors.  
\cref{sec:05} applies the proposed methodology to a masonry wall example, illustrating the quantification of variability reductions and resulting improvements in structural reliability.  
Finally, \cref{sec:06} concludes by summarizing the main findings, discussing broader implications, and identifying avenues for future research.

By bridging the gap between conformity assessment practices and semi-probabilistic design frameworks, this work provides a systematic, quantitative approach for incorporating quality control outcomes into structural reliability analyses. The proposed methodology not only enhances safety margins but also supports resource-efficient construction, offering a pathway for refining current design standards and ensuring sustainable, reliable built environments.
\section{State of the Art} \label{sec:02} Numerous studies emphasize the importance of quality control in ensuring structural reliability. While the positive impact of quality control is widely acknowledged, its quantitative assessment remains insufficiently addressed, particularly for materials beyond reinforced concrete.

The statistical treatment of uncertainty in structural reliability can be traced back to Galton, Pearson, and Neyman, who introduced the concept of confidence intervals to quantify uncertainty in observed data \cite{neymanOutlineTheoryStatistical1937}. In parallel, Shewhart and Deming developed systematic methods for industrial quality control, pioneering tools such as control charts and sampling techniques \cite{shewhartEconomicControlQuality1923}. These approaches laid the groundwork for integrating quality control principles into engineering applications.

Veneziano \cite{venezianoStatisticalEstimationPrediction1974} extended this foundation by highlighting the need to account for both statistical and model uncertainties to achieve accurate predictions of reliability. Rackwitz advanced this line of thought by applying Bayesian updating methods to refine distributions of material properties based on conformity assessments, enabling reliability estimates to evolve as new evidence emerged \cite{rackwitzQualitatsangebotBetonII1977, rackwitzUberWirkungAbnahmekontrollen1979, rackwitzDistributionQualitiesOffered1979, rackwitzPredictiveDistributionStrength1983}. This iterative approach connected quality control to uncertainty reduction in material parameters.

While operating characteristic (OC) curves and filter curves were already known concepts, their application to structural reliability was first recognized as significant by Rackwitz and Taerwe. Taerwe applied filter curves to describe quality control as a probabilistic mechanism that filters material distributions by rejecting non-conforming samples \cite{taerweGeneralBasisSelection1986}. Taerwe further emphasized the importance of autocorrelation in sequential sampling, demonstrating that neglecting temporal or spatial dependencies can distort reliability estimates \cite{taerweInfluenceAutocorrelationOClines1987, taerweSerialCorrelationConcrete1987, taerweEvaluationCompoundCompliance1988}. 

Additional contributions were made by Parkinson and Govindaraju, who optimized sampling plans and investigated their influence on quality control outcomes \cite{parkinsonControlOptimisationVariability1987, parkinsonOptimumSamplingPlans1988, parkinsonEffectsVariablesSamples1991, govindarajuSingleSamplingPlans1990}. Their work provided practical tools for balancing safety and economic efficiency. More recently, Caspeele and Taerwe extended Bayesian methods to evaluate conformity criteria for concrete materials, demonstrating how probabilistic approaches can improve structural durability assessments \cite{caspeeleProbabilisticEvaluationConformity2010, caspeeleProbabilisticEvaluationConformity2011, caspeeleBayesianAssessmentCharacteristic2012, caspeeleNumericalBayesianUpdating2013, caspeeleQualityControlStructural2014}.

Despite these advancements, significant gaps remain in current research. First, most studies focus predominantly on concrete materials, leaving other construction materials such as masonry, steel, and timber underexplored. Second, while addressing statistical uncertainties in structural reliability, previous work has primarily focused on the mean reliability index. Capturing the variance of the reliability index is essential for a comprehensive assessment under uncertainty, yet such a treatment requires a fundamental and detailed theoretical approach that lies outside the scope of this paper.

\section{Theoretical Framework} \label{sec:03}

\subsection{Conformity Assessment as a Filtering Mechanism}
Conformity assessment in construction functions as a probabilistic filter that removes non-compliant batches, thereby improving the quality of accepted materials. Integrating Bayesian statistics into this process enables the quantification of its filtering effect. By combining prior knowledge of material properties with new inspection data, it is possible to derive updated (posterior) distributions that reflect the improved quality after conformity assessment \cite{caspeeleNumericalBayesianUpdating2013}. Prior research has successfully applied Bayesian updating to assess how conformity controls enhance structural reliability \cite{rackwitzUberWirkungAbnahmekontrollen1979,rackwitzDistributionQualitiesOffered1979,rackwitzPredictiveDistributionStrength1983,caspeeleBayesianAssessmentCharacteristic2012,caspeeleProbabilisticEvaluationConformity2011}.

\subsection{Prior Information and Distributions}
Prior distributions $f_i(\mu, \sigma)$ represent initial estimates of relevant parameters, such as material strength, before quality testing occurs. These distributions rely on existing data sources, including historical measurements, expert judgment, literature surveys, and standards \cite{rackwitzPredictiveDistributionStrength1983,caspeeleNumericalBayesianUpdating2013}. \textit{Maximum Likelihood Estimation} (MLE) often serves as the starting point for deriving parameters from past observations. Alternatively, Bayesian updating can iteratively refine the distributions as new quality control data emerge, ensuring that the representation of material variability remains current and evidence-based.

When limited data are available or parametric assumptions are uncertain, non-parametric methods like bootstrap techniques can provide reliable estimates. In many applications, normal-gamma or lognormal-gamma distributions are preferred due to their conjugacy, which simplifies the Bayesian updating process. \cref{tab:hyperparameters} provides an example of hyperparameters for a normal-gamma distribution used in characterizing initial uncertainties of masonry materials and geometric parameters which will be used later in \cref{sec:05}.

\begin{table}[h!]
  \centering
  \caption{Hyperparameters of a Normal-Gamma Distribution for Masonry Mortar with a Compressive Strength of 5 MPa, Concrete Masonry Units, and Relative Eccentricity at Mid-Wall (Mean: 0.05). These values define the prior distributions before any quality assessment.}
  \label{tab:hyperparameters}
  \renewcommand{\arraystretch}{1.3}
  \setlength{\tabcolsep}{5pt}
  \begin{tabular}{p{4cm} p{3.5cm} p{2cm} p{2cm} p{2cm}}
      \hline
      \textbf{Parameter} & \textbf{Formula} & \textbf{Concrete Masonry Units} & \textbf{Masonry Mortar} & \textbf{Relative Eccentricity} \\ 
      \hline
      Prior Std. Deviation & $Q_0 = \sqrt{\ln(1+V_0^2)}$ & 0.177 & 0.198 & 0.340 \\
      Prior Mean & $\mu_0 = \ln(\bar{X}) - \tfrac{1}{2} Q_0^2$ & 2.692 & 1.590 & -3.054 \\
      Prior Precision & $\kappa_0 = n$ & 6 & 6 & 6 \\
      Shape Param. (Gamma) & $\alpha_0 = n / 2$ & 3 & 3 & 3 \\
      Rate Param. (Gamma) & $\beta_0 = \alpha_0 Q_0^2$ & 0.0957 & 0.118 & 0.347 \\
      \hline
  \end{tabular}
\end{table}

\subsection{Acceptance Functions, OC Curves, and Autocorrelation Effects}
\begin{figure}[h!]
  \centering
  \includegraphics[width=0.95\textwidth]{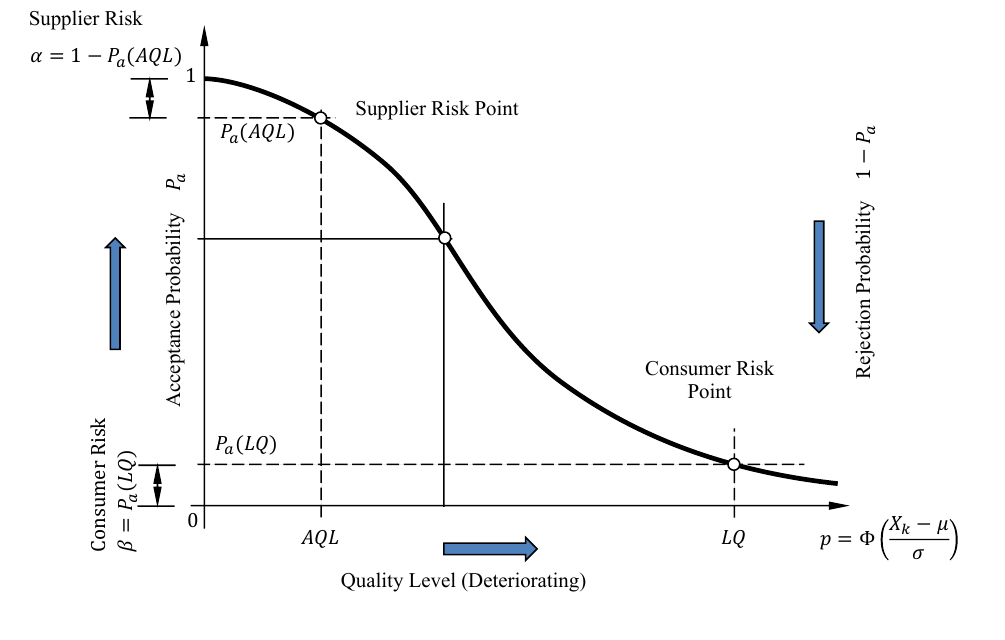}
  \caption{Example of an OC curve illustrating how acceptance probability ($P_a$) changes with varying levels of product quality ($p$). The acceptable quality limit (AQL) and limiting quality (LQ) indicate target thresholds that define supplier and consumer risks.}
  \label{fig:01}
\end{figure}

Operating characteristic (OC) curves, developed by Shewhart \cite{shewhartEconomicControlQuality1923} and refined by Dodge and Romig \cite{dodgeSamplingInspectionTables1959}, plot the probability of accepting a batch ($P_a$) against its defect rate ($p$). They serve as fundamental tools for visualizing and quantifying quality control performance. OC curves inform decisions about acceptable risk levels for both suppliers (who risk rejecting good products) and consumers (who risk accepting poor-quality products) \cite{DINISO28591,montgomeryIntroductionStatisticalQuality2009} \cref{fig:01}.

In construction, OC curves help translate conformity criteria into \textit{probabilistic acceptance functions}, relating statistical parameters (e.g., mean and standard deviation of a property) to acceptance probabilities. Although standards like DIN EN 206 specify conformity requirements for concrete, they do not explicitly provide OC curves \cite{DIN206202106}. Defining such curves for a broader range of construction materials would improve the understanding of quality control outcomes on structural reliability.

\begin{figure}[h!]
  \centering
  \includegraphics[width=0.95\textwidth]{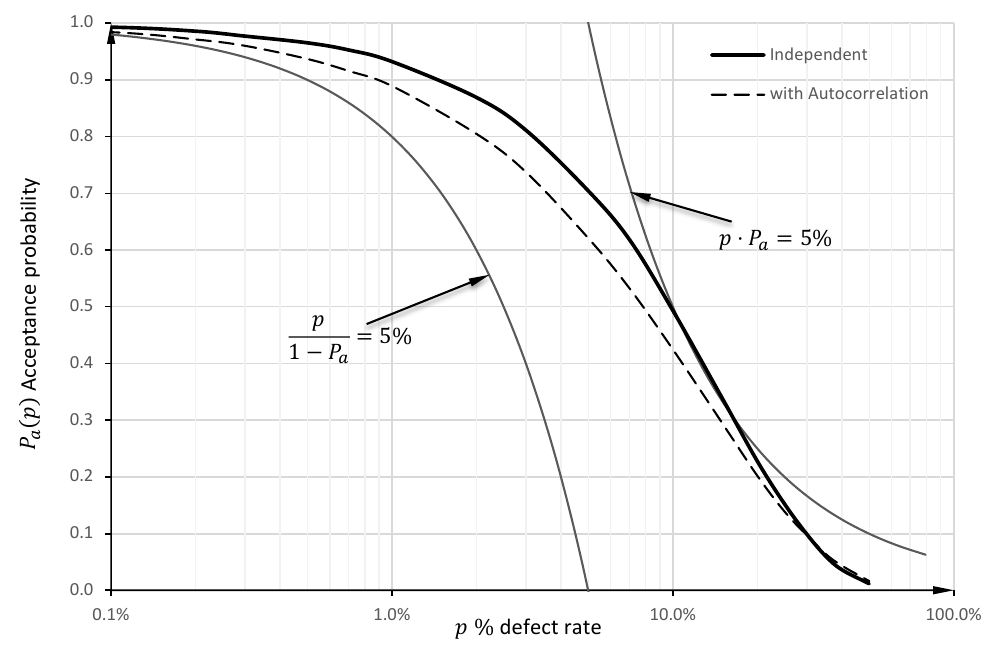}
  \caption{OC curves for the mean compressive strength of concrete masonry units, comparing independent samples and autocorrelated cases. Accounting for autocorrelation provides more realistic acceptance probabilities.}
  \label{fig:02}
\end{figure}

Autocorrelation, or the correlation between successive samples, further complicates the interpretation of OC curves. Environmental factors, production consistency, or process adjustments can introduce correlations that alter the variability of observed quality measures. Ignoring autocorrelation may lead to overoptimistic acceptance probabilities, as it downplays the impact of correlated fluctuations in quality \cite{taerweInfluenceAutocorrelationOClines1987} \cref{fig:02}.

Acceptance functions $P_a(\mu, \sigma)$ derive from OC curves and represent the probability that a batch with mean $\mu$ and standard deviation $\sigma$ meets conformity criteria:
\begin{equation}
    P_a(\mu, \sigma) = P(\text{batch accepted} \mid \mu, \sigma).
\end{equation}
These functions form the link between the statistical representation of material properties and the binary decision of acceptance or rejection.

\subsection{Bayesian Updating}
Bayesian updating integrates prior distributions and acceptance functions to yield posterior distributions that reflect the improved quality after conformity assessment:
\begin{equation}
    f_o(\mu, \sigma) = \frac{P_a(\mu, \sigma) f_i(\mu, \sigma)}{\iint P_a(\mu, \sigma) f_i(\mu, \sigma) \, d\mu \, d\sigma}.
    \label{eq:posterior_distribution}
\end{equation}

Here, $f_i(\mu, \sigma)$ represents the prior distribution of material parameters. Direct integration can be challenging; thus, Markov Chain Monte Carlo (MCMC) methods, such as the Metropolis-Hastings algorithm \cite{hastingsMonteCarloSampling1970}, are often employed to approximate the posterior distribution, especially in complex, high-dimensional problems.

After determining the posterior distribution $f_o(\mu, \sigma)$, the predictive distribution $g_o(X)$ for a lognormally distributed property $X$ can be computed by integrating out the parameters:
\begin{equation}
    g_o(X) \propto \iint \frac{1}{\sigma_{\ln X}} \, \phi\left(\frac{\ln X - \mu_{\ln X}}{\sigma_{\ln X}}\right) f_o(\mu_{\ln X}, \sigma_{\ln X}) \, d\sigma \, d\mu,
    \label{eq:predictive_distribution}
\end{equation}
where $\phi$ is the standard normal density. The predictive distribution provides an updated estimate of expected material properties, reflecting improvements achieved through conformity assessment.

\begin{figure}[h!]
  \centering
  \includegraphics[width=\textwidth]{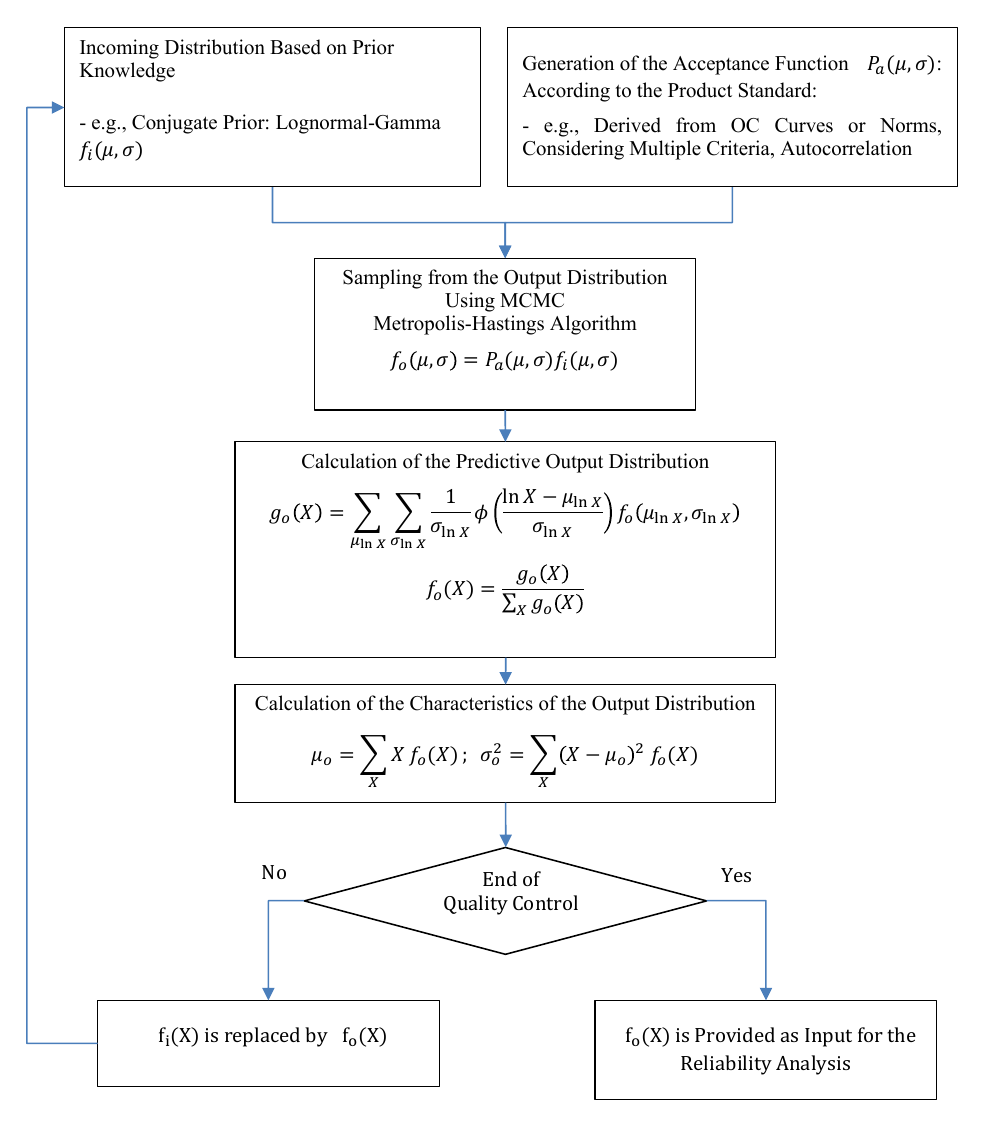}
  \caption{Flowchart illustrating the Bayesian updating process. Starting from a prior distribution and an acceptance function derived from OC curves, MCMC sampling produces the posterior distribution. The predictive distribution is then computed to assess the refined quality of the materials.}
  \label{fig:03}

\end{figure}

\section{Integrating Quality Control and Partial Safety Factors} \label{sec:04}

Reliability-based design methods in structural engineering often employ partial safety factors to balance safety and cost considerations. These partial factors depend on the statistical variability of input parameters, commonly represented by the coefficient of variation \(V_X = \sigma_X/\mu_X\), where \(\sigma_X\) and \(\mu_X\) are the standard deviation and mean of the parameter \(X\), respectively. Higher values of \(V_X\) indicate greater uncertainty and typically require larger partial safety factors to maintain the desired level of reliability. This section introduces a practical approach for incorporating the effects of quality control measures into the calibration of partial safety factors, ensuring that improvements in material quality translate into tangible safety and economic benefits.

\subsection{Derivation of Improvement Factors}

Quality control measures can reduce parameter variability, thereby enhancing structural reliability. To quantify the resulting improvements in partial safety factors, the improvement factor \(r\) is introduced. Consider a lognormally distributed parameter \(X\). The partial safety factors before and after quality control can be expressed as:
\begin{equation}
    \gamma_{\text{in}} = b \exp\bigl((\alpha_X \beta - k) V_{\text{in}}\bigr), \quad \gamma_{\text{out}} = b \exp\bigl((\alpha_X \beta - k) V_{\text{out}}\bigr),
    \label{eq:safety_factor_before_after}
\end{equation}
where \(b\) denotes the model bias, \(\alpha_X\) the resistance sensitivity, \(\beta\) the target reliability index, and \(k\) a fractile factor (e.g., \(k = 1.645\) for the 5\% fractile).

The improvement factor \(r\) is defined as the ratio between the partial safety factors before and after quality control:
\begin{equation}
    r = \frac{\gamma_{\text{in}}}{\gamma_{\text{out}}} = \exp\bigl((\alpha_X \beta - k)(V_{\text{in}} - V_{\text{out}})\bigr).
    \label{eq:improvement_factor2}
\end{equation}

For typical values \(\alpha_X = 0.8\) and \(\beta = 3.8\), the expression \((\alpha_X \beta - k)\) approximately equals 1.4. Thus:
\begin{equation}
    r = \exp\bigl(1.4(V_{\text{in}} - V_{\text{out}})\bigr) = \exp(1.4 \Delta V),
    \label{eq:improvement_factor_simplified3}
\end{equation}
where \(\Delta V = V_{\text{in}} - V_{\text{out}}\) represents the reduction in the coefficient of variation due to quality control. This relationship provides a straightforward means of quantifying the impact of multiple quality control measures, as the total reduction \(\Delta V\) is simply the sum of individual reductions. This cumulative approach ensures that all quality control initiatives can be collectively accounted for.

\subsection{Influence of Quality Control on Partial Safety Factors in Structural Systems} \label{improvement_factor}

In structural systems composed of multiple parameters, the improvement in partial safety factors must consider the combined effect of all relevant variables. According to Bakeer \cite{bakeerGeneralPartialSafety2023}, the degree of homogeneity \(n_{X_i}\) of a parameter \(X_i\) describes how changes in an input parameter affect changes in the overall system response:
\begin{equation}
    n_{X_i} = \frac{\text{relative change in system output}}{\text{relative change in system input } X_i}.
    \label{eq:homogeneity_degree}
\end{equation}

For a resistance model \(R = R(X)\), the homogeneity degree can be expressed as:
\begin{equation}
    n = \frac{X_d}{R_d}\frac{dR(R_d)}{dX} \approx \frac{(\Delta R / R_d)}{(\Delta X / X_d)},
    \label{eq:resistance_homogeneity}
\end{equation}
where \(R_d\) and \(X_d\) are the design values of the resistance and the input parameter, respectively.

If the resistance model involves multiple independent parameters, it can be approximated in logarithmic form:
\begin{equation}
    \log R = \sum_{i=1}^N n_i \log X_i,
    \label{eq:log_resistance_model}
\end{equation}
where \(n_i\) represents the homogeneity degree for each parameter \(X_i\).

For lognormally distributed parameters, the logarithmic standard deviations \(Q_R\) and \(Q_i\) link coefficients of variation to resistance variability:
\begin{equation}
    Q_R = \sqrt{\ln(1+V_R^2)}, \quad Q_i = \sqrt{\ln(1+V_i^2)}.
    \label{eq:log_std_dev_input}
\end{equation}

The aggregated variance of resistance \(Q_R^2\) is the sum of the weighted variances of each parameter:
\begin{equation}
    Q_R^2 = \sum_{i=1}^N n_i^2 Q_i^2.
    \label{eq:aggregated_variance}
\end{equation}

The partial safety factor for the resistance then reads:
\begin{equation}
    \gamma_R = \exp\bigl((\alpha_R \beta - k)Q_R\bigr),
    \label{eq:partial_safety_factor}
\end{equation}
where \(\alpha_R\) is the resistance sensitivity factor and \(\beta\) the target reliability index.

Similar to the single-parameter case, an improvement factor \(r\) can quantify changes in \(\gamma_R\) before and after quality control:
\begin{equation}
    r = \exp\bigl((\alpha_R \beta - k)(Q_{R,\text{in}} - Q_{R,\text{out}})\bigr).
    \label{eq:improvement_factor4}
\end{equation}

\begin{figure}[h!]
  \centering
  \includegraphics[width=0.9\textwidth]{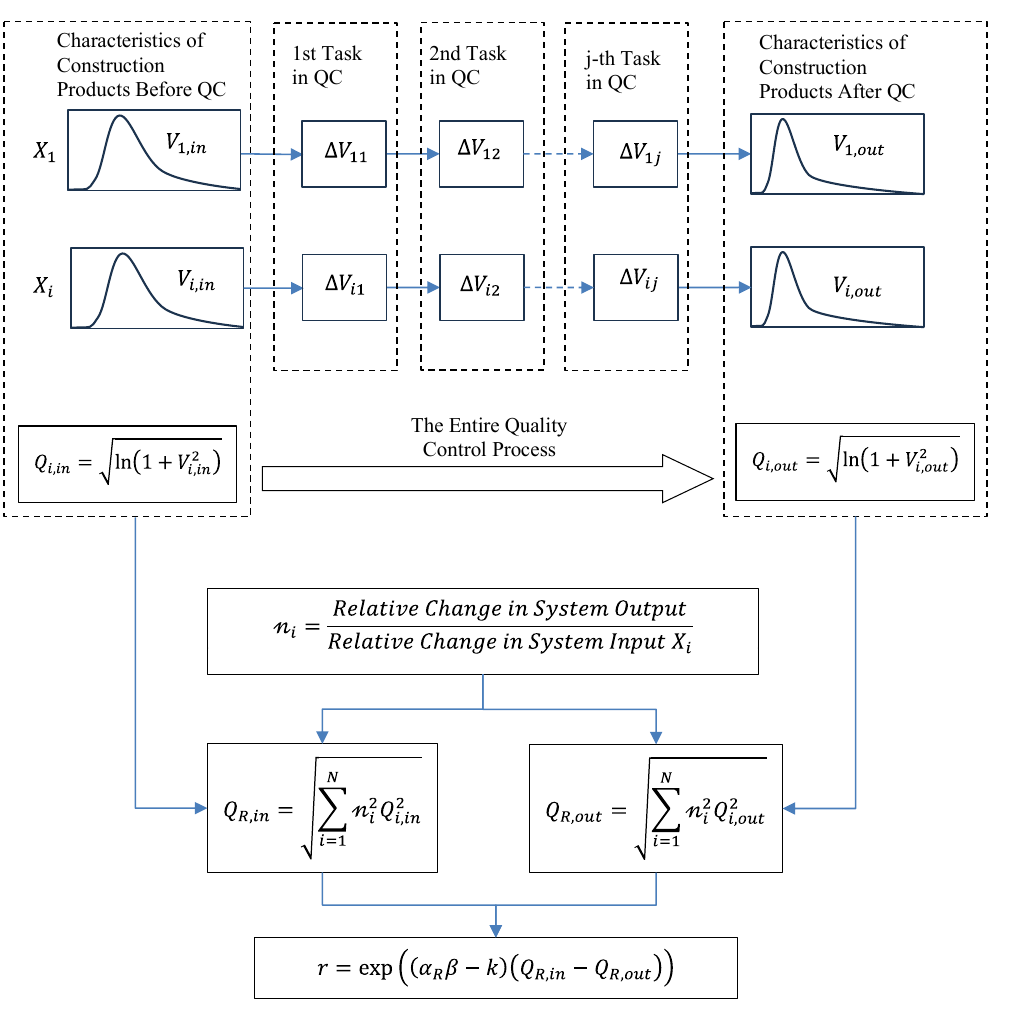}
  \caption{Flowchart illustrating the process of quantifying the effects of quality control measures on partial safety factors. After determining the reductions in parameter variability, the improvement factors can be calculated and applied to calibrate partial safety factors.}
  \label{fig:04}
\end{figure}

This framework \cref{fig:04} provides a consistent and clear method to quantify how systematic quality control measures influence partial safety factors at both the material and system levels. By directly relating improvements in material quality (as reflected by reduced variability) to reduced partial safety factors, the approach supports the development of more reliable and cost-effective design solutions.

\section{Application Example} \label{sec:05}

\subsection{Model Description and Parameter Sensitivity}

This example examines how quality control measures affect the reliability of a vertically loaded masonry wall, following E DIN EN 1996-1-1:2019-09 \cite{DIN199611201909} and using background information from \cite{bakeerEmpiricalEstimationLoad2016,bakeerBucklingMasonryWalls2017}. The resistance per unit length, \( R \), is expressed as:
\begin{equation}
    R = f \cdot \Phi \cdot t,
    \label{eq:resistance_per_length}
\end{equation}
where \( f \) is the compressive strength of the masonry, \( \Phi \) is a reduction factor defined in Annex G of the standard \cite{DIN199611201909}, and \( t \) is the wall thickness. The factor \(\Phi\) depends on the slenderness ratio \(\lambda\) and the relative eccentricity \(r_e\):
\begin{equation}
    \Phi =
    \begin{cases} 
        A - \frac{\lambda^2}{2.58A}, & \text{if } \lambda < 1.14A, \\
        \frac{0.65A^3}{\lambda^2}, & \text{if } \lambda \geq 1.14A,
    \end{cases}
    \label{eq:reduction_factor}
\end{equation}
where \( A = 1 - 2r_e \) and \( r_e = e/t \).

The slenderness ratio \(\lambda\) is given by:
\begin{equation}
    \lambda = \frac{h}{t} \sqrt{\frac{f}{E}},
    \label{eq:slenderness_ratio}
\end{equation}
where the modulus of elasticity \( E \) is defined as \( E = K_E \cdot f \).

\begin{figure}[h!]
    \centering
    \includegraphics[width=0.6\textwidth]{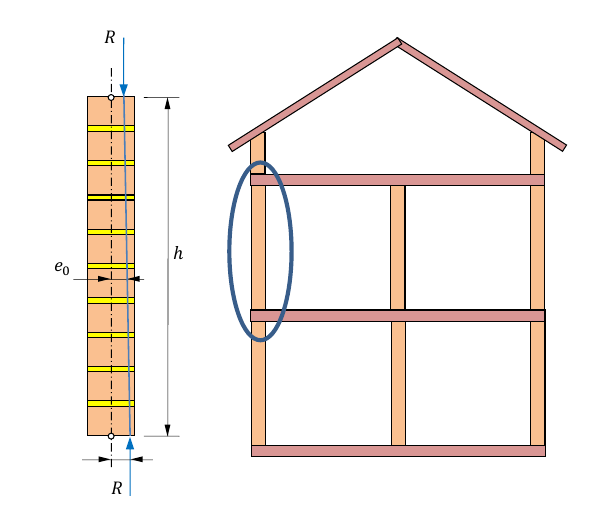}
    \caption{Schematic representation of a vertically loaded masonry wall. The resistance is influenced by material properties, such as the compressive strength of masonry and mortar, and geometric imperfections, such as initial eccentricity. The parameters used in this analysis are: wall height \( h = 3.3 \, \mathrm{m} \), wall thickness \( t = 240 \, \mathrm{mm} \), masonry unit compressive strength \( f_b = 15 \, \mathrm{MPa} \), mortar compressive strength \( f_m = 5 \, \mathrm{MPa} \), and eccentricity \( e = 24 \, \mathrm{mm} \). Additional coefficients include \( \alpha = 0.585 \), \( \beta = 0.162 \), \( K = 0.79 \), and \( K_E = 2400 \).}
    \label{fig:05}
\end{figure}

The wall considered here is constructed using concrete masonry units in accordance with DIN EN 771-3:2015-11 \cite{DIN7712201511} and mortar M5 with \( f_m = 5 \, \mathrm{N/mm^2} \) \cite{DIN9982201702}. The characteristic compressive strength of the masonry \( f_k \) is:
\begin{equation}
    f_k = K \cdot f_b^\alpha \cdot f_m^\beta = 5.0 \, \mathrm{N/mm^2},
    \label{eq:compressive_strength}
\end{equation}
with \( f_b = 15 \, \mathrm{N/mm^2} \), \( K = 0.79 \), \(\alpha = 0.585\), \(\beta = 0.162\) \cite{DIN199611NA}. The correlation factor \(K_E = 2400\), and the partial safety factor \(\gamma_M = 1.5\).

This example focuses on three parameters: (1) Compressive strength of the mortar, (2) Compressive strength of the masonry unit, and (3) Eccentricity of initial imperfections. Each parameter is subject to quality control measures aimed at reducing its variability.

\begin{figure}[h!]
    \centering
    \includegraphics[width=0.8\textwidth]{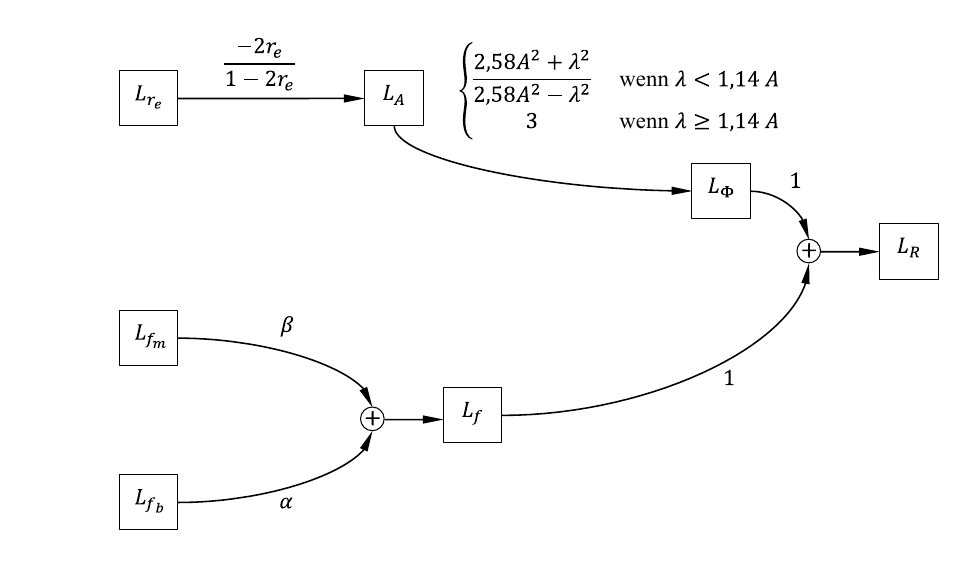}
    \caption{Sensitivity flow diagram indicating the relative influence of input parameters on the wall's resistance.}

    \label{fig:06}
\end{figure}

Figure \ref{fig:06} shows a sensitivity flow diagram that indicates how changes in input parameters affect the structural resistance. 

For the given example:   $n_{f_b} = \alpha = 0.585, \quad n_{f_m} = \beta = 0.162$. The slenderness ratio \( r_h \) and parameter \(\lambda\) are computed as:     $r_h = h/t = 13.75$, $\lambda = 0.281$. The relative eccentricity is: $r_e  = 0.1$, and $A = 1 - 2r_e = 0.8$.

Since \( \lambda = 0.281 < 1.14A = 0.912 \), the homogeneity degree for eccentricity \( n_{r_e} \) is:
\begin{equation}
    n_{r_e} = \frac{2r_e}{A} \cdot \frac{2.58A^2 + \lambda^2}{2.58A^2 - \lambda^2} = 0.275.
\end{equation}

These homogeneity degrees allow the relative change in resistance \( L_R \) to be expressed as:
\begin{equation}
    L_R = 0.585 \cdot L_{f_b} + 0.162 \cdot L_{f_m} + 0.275 \cdot L_{r_e},
    \label{eq:resistance_change}
\end{equation}
quantifying how variations in material strength and execution quality affect the overall structural resistance.

\subsection{Evaluating Quality Control with OC Curves}

Operating Characteristic (OC) curves quantify how quality control measures influence acceptance probabilities and, consequently, structural reliability. This evaluation considers three aspects relevant to the masonry wall: concrete masonry units, masonry mortar, and execution quality.

\subsubsection{Concrete Masonry Units}

Quality control for concrete masonry units follows DIN EN 771-3:2015-11 \cite{DIN7712201511}. Testing involves characteristic (\( f_c \)) and mean (\( f_m \)) compressive strengths. If initial samples fail to meet the specified criteria, a second sample set is tested. Acceptance occurs only if both sets satisfy the requirements.

By applying the given acceptance criteria, OC curves were generated using the characteristic and mean strength values. These curves relate acceptance probability to defect rate, indicating that basing quality control on the characteristic value \( f_c \) achieves a better balance between safety and cost-efficiency than focusing solely on \( f_m \).

To incorporate sample dependencies, an AR(2) model was used:
\begin{equation}
    x_k = 0.4\, x_{k-1} + 0.2\, x_{k-2} + N(0.4\mu_x,0.8\sigma_x^2).
\end{equation}

The analysis revealed that autocorrelated OC curves have higher acceptance probabilities at low defect rates and lower acceptance probabilities at high defect rates, which improves the balance between safety and cost-efficiency.

\subsubsection{Masonry Mortar}

Unlike masonry units, masonry mortar lacks explicit conformity criteria. Drawing on DIN EN 206:2021-06 \cite{DIN206202106}, the following criterion was adopted:
\begin{equation}
    \bar{x} > x_k + 1.48 \cdot s,
\end{equation}
where \(\bar{x}\) is the mean value, \(x_k\) is the characteristic value, and \(s\) is the standard deviation of compressive strength. 

\subsubsection{Execution Control}

Execution control aims to minimize initial imperfections represented by the relative eccentricity \( r_e \). The masonry execution is accepted if ten randomly selected measurements do not exceed an eccentricity of 5\%. The OC curve for execution control showed a significant reduction in second-order effects, thereby increasing the load-bearing capacity of the wall.

\subsection{Numerical Simulation}

A numerical simulation combined the prior distributions, acceptance criteria, and OC curves to determine how quality control affected material variability. The coefficient of variation \( V = \sigma/\mu \), a key parameter in the calibration of partial safety factors, was monitored before and after successive quality control interventions.

Results showed a progressive decrease in \( V \) following each stage of quality control. Incorporating autocorrelation and implementing multiple checks led to a more concentrated distribution of material properties around the expected value, reflecting a notable improvement in quality.

\subsection{Results of Quality Control}

Applying systematic quality control measures to masonry units, mortar, and execution quality led to substantial reductions in the coefficient of variation \( V \), demonstrating improved material consistency and enhanced structural reliability. The observed decreases in \( V \) for each parameter were:

\begin{itemize}
    \item \textbf{Concrete Masonry Units}: \(V\) decreased from 0.25 to 0.20 after the first quality control, and then to 0.18 after the second stage.
    \item \textbf{Masonry Mortar}: \(V\) reduced from 0.27 to 0.22 after the first control, and further to 0.20 after the second control.
    \item \textbf{Execution Quality}: \(V\) declined from 0.47 to 0.38 after the first control, and then to 0.34 after the second control.
\end{itemize}

\begin{figure}[h!]
    \centering
    \includegraphics[width=\textwidth]{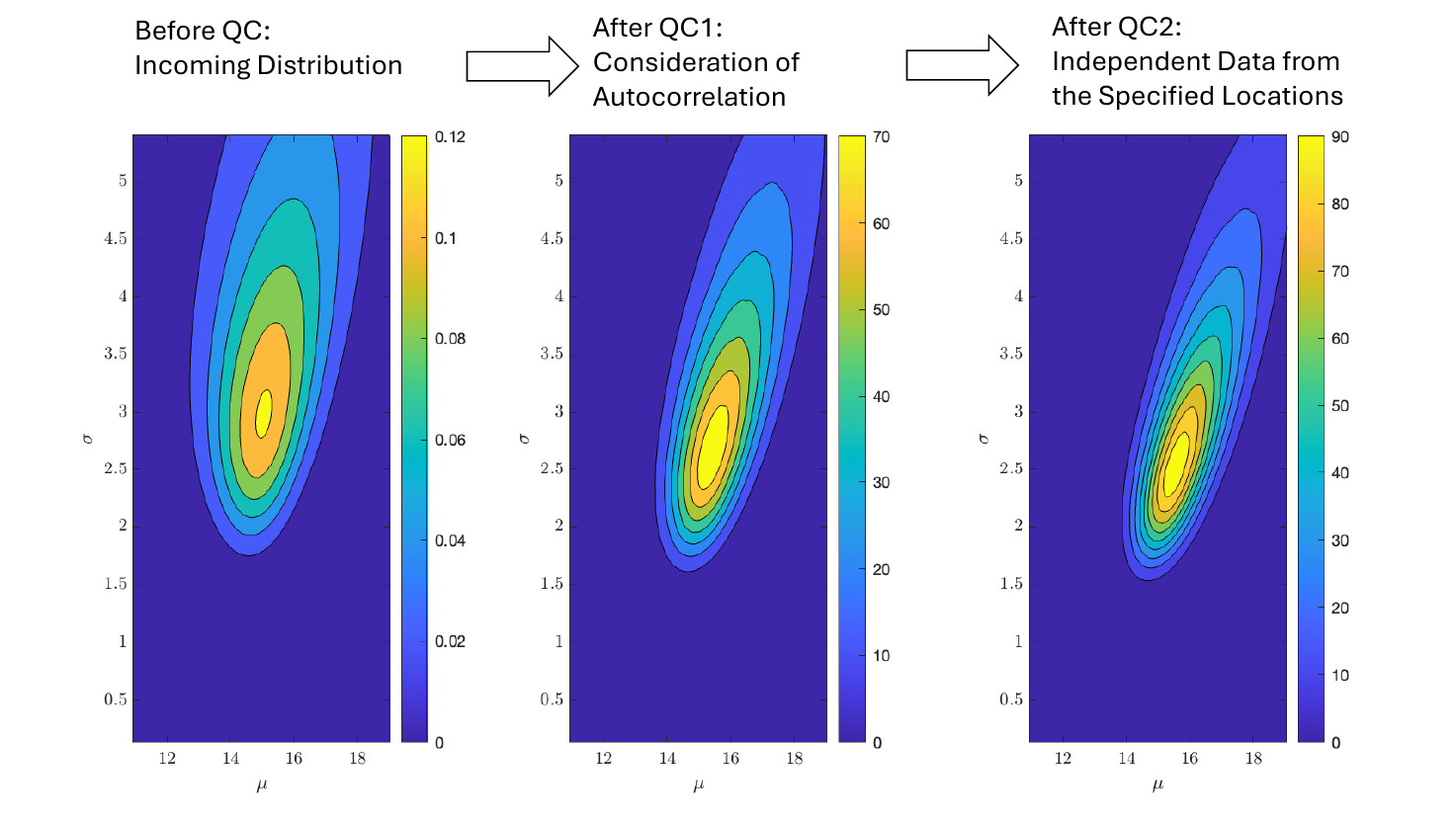} 
    \caption{Comparison of the distributions of the compressive strength of masonry unit before and after the first and second stage of quality control. Incoming distribution (left): This distribution is based on the previously calculated hyperparameters of the normal-gamma distribution and represents the uncertainty regarding the mean $\mu$ and the standard deviation $\sigma$ of the compressive strength before any quality assurance measures. The first Outgoing distribution (Middle): After the first stage of quality control. The second Outgoing distribution (right):  After the second stage of quality control.}
    \label{fig:07}
  \end{figure}

\begin{figure}[h!]
    \centering
    \includegraphics[width=\textwidth]{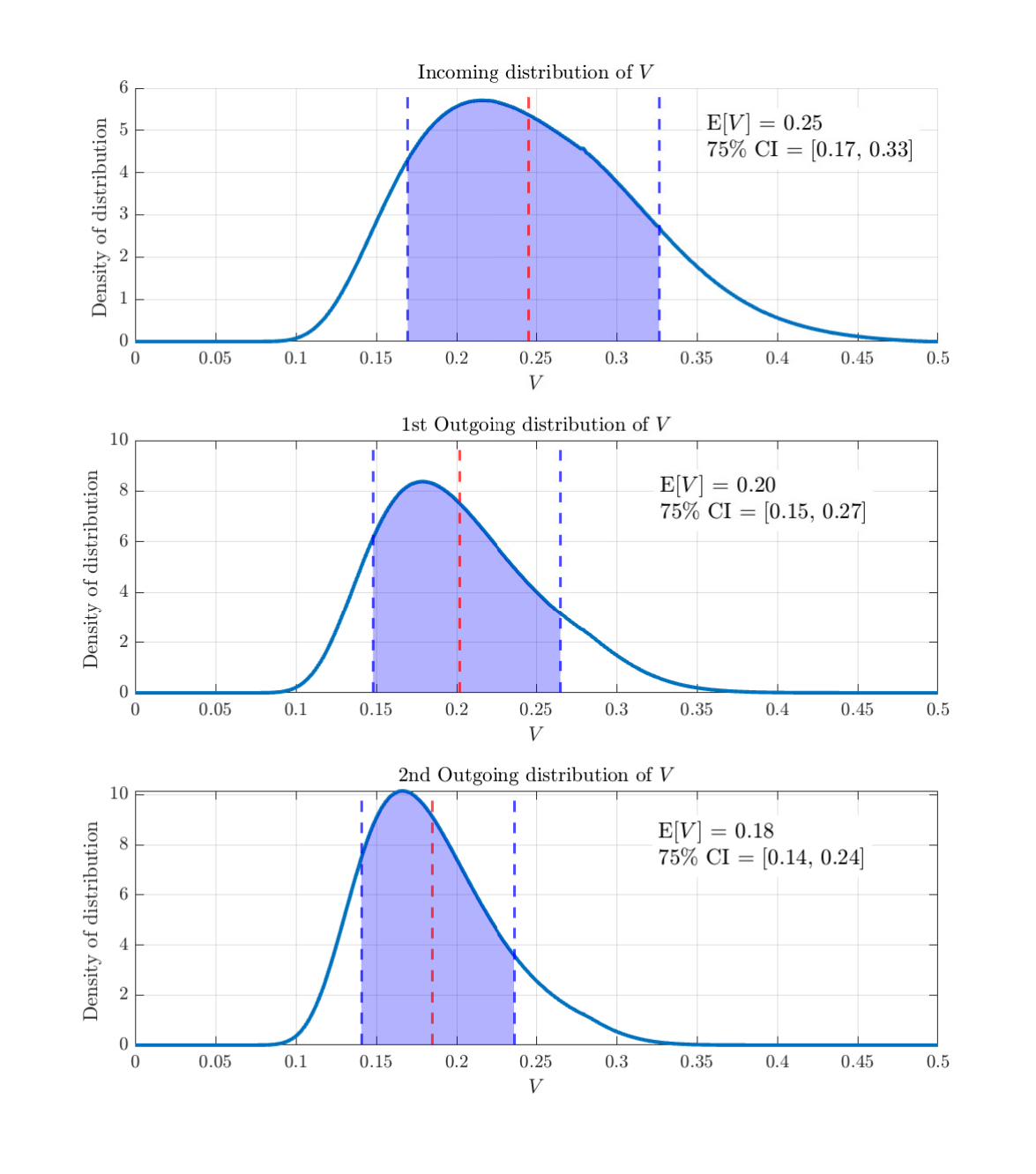} 
    \caption{Distribution of the coefficient of variation \( V \) of the compressive strength of concrete masonry units before and after quality control (75\% credibility interval). 
    Top: Distribution of \( V \) before quality control. 
    Middle: Distribution of \( V \) after the first quality control. 
    Bottom: Distribution of \( V \) after the second quality control.}
    \label{fig:08}
  \end{figure}

Figures~\ref{fig:07} and \ref{fig:08} illustrate the progressive refinement of the distributions as additional quality control steps are introduced. The resulting, more uniform material properties serve as a quantitative foundation for recalibrating partial safety factors. This approach aligns with the semi-probabilistic design framework, offering a systematic method to incorporate quality control outcomes into reliability-based design.

These improvements highlight that certain parameters contribute more significantly to overall safety gains. In particular, controlling masonry units and execution quality yields notable reductions in the partial safety factor for resistance. As summarized in Table~\ref{tab:02}, when all three parameters (masonry units, mortar, execution) are subjected to quality control, the partial safety factor decreases from the baseline value of 1.5 to as low as 1.40, representing approximately an 8\% improvement. Table~\ref{tab:03} further illustrates the range of improvement factors and corresponding partial safety factors, considering both the expected values of \( V \) and the upper limits of the 75\% credibility intervals.

The data indicate that quality control of mortar alone provides negligible safety gains for the current example. By contrast, quality control measures focused on masonry units and execution yield meaningful reductions in uncertainty and enhanced reliability.

\renewcommand{\arraystretch}{1.3} 
\begin{table}[h!]
\centering
\caption{Determination of the safety improvement based on the partial safety factor after quality control. The following components have undergone quality control: Bricks, mortar, execution.}
\label{tab:02}
\begin{tabular}{p{5cm} p{2cm} p{2cm} p{2cm} p{2cm}}
\toprule
\textbf{Quality-Controlled Component} & \( n \) & \textbf{Incoming} & \textbf{1st Outgoing} & \textbf{2nd Outgoing} \\ \midrule
Masonry Unit                 & 0.585  & 0.250             & 0.200                 & 0.180                 \\
Mortar                       & 0.162  & 0.270             & 0.220                 & 0.200                 \\
Execution                    & 0.275  & 0.470             & 0.380                 & 0.340                 \\
Model                        & 1.000  & 0.050             & \textit{No QC}        & \textit{No QC}        \\ 
\midrule
\multicolumn{5}{l}{\textbf{Coefficient of Variation for Resistance} \( Q_R \)} \\ \midrule
\textbf{\( Q_R = \)}                  &        &0.200              & 0.165                 & 0.151                \\ 
\midrule
\multicolumn{5}{l}{\textbf{Improvement Factors and Partial Safety Factor}} \\ \midrule
 &      &  & \textbf{1st QC} & \textbf{2nd QC} \\
 \multicolumn{3}{l}{\( \Delta Q_R = Q_{R,\text{in}} - Q_{R,\text{out}} \) }               & 0.035             & 0.050          \\ 
\rowcolor[HTML]{E5E5E5} 
\multicolumn{3}{l}{{Improvement Factor \( r = \exp{((\alpha_R \beta - k) \Delta Q_R)} \)} }         & 1.05              & 1.07     \\ 
\rowcolor[HTML]{E5E5E5} 
\multicolumn{3}{l}{{Improved Partial Safety Factor \( \gamma_M = \)}}       & 1.43              & 1.40                   \\

\bottomrule
\end{tabular}
\end{table}

\begin{table}[h!]
  \centering
  \caption{Improvement factor \( r \) and improved partial safety factor \( \gamma_M \). 
  (1) Based on the expected value of the coefficient of variation, 
  (2) based on the coefficient of variation at the upper limit of the 75\% credibility interval.}
  \label{tab:03}
  \small 
  \renewcommand{\arraystretch}{1.3} 
  \setlength{\tabcolsep}{5pt} 
  \begin{tabularx}{\textwidth}{>{\raggedright\arraybackslash}X 
                                >{\centering\arraybackslash}m{3.3cm}
                                >{\centering\arraybackslash}m{2cm}
                                >{\centering\arraybackslash}m{2cm}
                                >{\centering\arraybackslash}m{2cm}}
      \toprule
      \textbf{Quality Control Task} & 
      \makecell{\textbf{Improvement} \\ \textbf{Factor} \( r \) \textbf{(1)}} & 
      \makecell{\textbf{Improved} \\ \textbf{PSF} \( \gamma_M \) \textbf{(1)}} & 
      \makecell{\textbf{Improvement} \\ \textbf{Factor} \( r \) \textbf{(2)}} & 
      \makecell{\textbf{Improved} \\\textbf{PSF} \( \gamma_M \) \textbf{(2)}} \\
      \midrule
      Masonry Units and Execution 1.QC        & 1.05 & 1.43 & 1.05 & 1.42 \\
      Masonry Units and Execution 1.QC + 2.QC  & 1.07 & 1.40 & 1.09 & 1.38 \\
      Masonry Units 1.QC                      & 1.03 & 1.46 & 1.03 & 1.46 \\
      Masonry Units 1.QC + 2.QC               & 1.04 & 1.45 & 1.05 & 1.43 \\
      Execution 1.QC                          & 1.02 & 1.47 & 1.02 & 1.47 \\
      Execution 1.QC + 2.QC                   & 1.03 & 1.46 & 1.03 & 1.45 \\
      \bottomrule
  \end{tabularx}
\end{table}

\section{Conclusion} \label{sec:06}

This investigation demonstrates that integrating quality control measures into reliability-based design methodologies can significantly enhance structural safety while optimizing resource allocation. By modeling conformity assessment as a probabilistic filter, applying Bayesian updating, and employing OC curves, the analysis quantifies how improved material and execution consistency reduce the coefficient of variation and, consequently, the partial safety factors required to maintain target reliability indices.

The results show that controlling parameters with higher importance leads to substantial reductions in partial safety factors. For the masonry wall considered in this study, systematic quality control lowered the partial safety factor from the baseline of 1.5 to 1.38 when multiple controls were applied. This corresponds to an improvement factor of about 1.09 and suggests potential material savings of around 8\%. In contrast, controlling parameters with a negligible impact, such as mortar properties in this example, is not cost-effective and does not significantly enhance safety.

These insights provide clear guidelines for practitioners and standardization bodies. By prioritizing quality control efforts on parameters that yield notable safety improvements and integrating these results into a semi-probabilistic design framework, the construction industry can achieve more reliable and sustainable structures. The methodology developed here—linking acceptance sampling, Bayesian updating, and importance factors—offers a practical, standards-compatible approach for systematically incorporating quality control outcomes into the calibration of partial safety factors.

The findings encourage future enhancements in building standards to include parameter-specific guidelines and OC curves, enabling more transparent and efficient evaluation of quality control practices. This approach supports the ongoing refinement of European norms, ensuring that investments in quality assurance directly translate into measurable safety gains and resource efficiency.

\section*{Acknowledgments}

This research was conducted within the framework of the project \textit{“Quality Control and Structural Reliability in Structural Engineering – B: Effects of the AVCP System in Production Control and integration with the Partial Safety Factor Concept”}, funded by the (German Institute for Structural Engineering) \textit{Deutsches Institut für Bautechnik} (DIBt).

The project was carried out at the \textit{Planungs- und Ingenieurbüro für Bauwesen} , Wichernstraße 12, 01445 Radebeul. Special thanks go to Dipl.-Ing. (FH) Anke Eis for her support to the research project.

\printbibliography
\end{document}